\documentstyle[aps,epsf]{revtex}  

\begin{document}        

\baselineskip 14pt

\title{Search for Long-lived Charged Massive Particles at CDF}
\author{Amy Connolly}
\author{For the CDF Collaboration}
\address{Lawrence Berkeley National Laboratory, One Cyclotron Road,
Berkeley, CA  94720, USA}
\address{Email:  connolly@cdfsg6.lbl.gov}  
%
\maketitle              

\begin{abstract}        
A search for long-lived charged massive particles 
in CDF's Run1b data sample is presented.  
The search looks for highly ionizing tracks which would result from slowly
moving massive particles.
We search for strongly
produced particles using a stable color triplet quark as
a reference model, and a separate search was performed
for weakly produced particles using long-lived sleptons
in Gauge-Mediated Supersymmetry Breaking as a reference model.  
No excess over background was observed, and we derive 
limits on the cross-sections
for production of these particles.   Prospects for RunII
are also discussed.

\end{abstract}   	

\section{Introduction}               

Many models for new physics include particles which may live
long enough to traverse our detector.  
 Such a 
particle may be long-lived due to a new conserved quantum number
(such as R-parity in SUSY), or its decays may be
suppressed.
If these
particles are charged, they could be detected
directly~\cite{theories,phenom,penet}.

The most stringest limits from direct searches 
for such particles come from LEPII and CDF.
LEPII has set mass limits of about 82 GeV/c$^{2}$ on stable
sleptons within the framework of Gauge-Mediated 
Supersymmetry Breaking (GMSB)~\cite{lep2}.
CDF set a limit of M $>$ 139 GeV/c$^{2}$ on stable quarks
from data taken in 88/89~\cite{cdf_run1}.  Other searches have also been
performed at LEP and other accelerators~\cite{other_searches}.

We have searched for Charged Massive Particles (CHAMPS) in 90 pb$^{-1}$
of Run1B data at CDF.  We have aimed to be as model independent
as possible, but the search naturally divides into two separate
searches, one for strongly produced particles and one for those
produced via the weak interaction.

The strongly produced particles
would have a larger production cross section,
so the region of
interest is at high mass where the background is expected to be low.
We expect
that these particles would fragment into an integer charged meson
within a jet.  For the strong search, we use a 4$^{th}$ generation
quark as a reference model.

Weakly interacting CHAMPS would have a lower production cross-section
so the region of sensitivity is at lower mass where the
background is higher.  These
weakly produced particles are expected to be isolated, which
allows us to reduce the background significantly.  We use Drell-Yan 
production of a slepton within a GMSB scenario as a reference model.


\section{Signature}

Massive particles will be
produced with relatively low $\beta\gamma$, 
so the signature is a highly ionizing track.
  We use the Central
Tracking Chamber (CTC) and the Silicon Vertex Detector (SVX) to
independently measure dE/dx.  For both
searches, we make a cut of $\beta\gamma <$ 0.85, and for the strong
search we additionally look at $\beta\gamma <$ 0.70 for added
sensitivity at low mass.  We searched
track by track from events which came in on three different triggers --
the muon trigger, $E_{T}\hspace{-0.17in}/\hspace{0.1in}> $ 
35 GeV trigger, and the
electron trigger.  The muon trigger 
is directly sensitive to a CHAMP since massive particles 
would be penetrating and appear as muons,
while all three triggers are sensitive to production of sleptons
in
cascade decays of other sparticles which produce in addition a neutrino,
electron or muon.




\section{Kinematic Cuts}

All considered tracks are
required to pass quality cuts which reduce backgrounds from misreconstructed
tracks.  Due to timing
considerations,
particles moving slower than $\beta\gamma$ = 0.4 are not reconstructed. For a
particle of mass 90 GeV/c$^{2}$, this corresponds to a minimum reconstructable
momentum of 35 GeV.  
We therefore require that
each track has a momentum greater than 35 GeV/c, which eliminates
much of our low momentum background.  We also cut on the mass 
$M_{SVX}$
calculated from the
dE/dx measurement in the SVX and momentum of the track.
For the strong search, we perform a search in bins of mass M,
and make a cut on $M_{SVX} > 0.6 \times M$.  Our background
falls off with momentum, and this $M_{SVX}$ cut forces a stiffer
dE/dx cut at low momentum where the background is higher.     
For the weak search, the mass cut is simply $M_{SVX} > 60 GeV/c^{2}$.
These tracks must additionally pass an isolation
cut, namely we require less than 4 GeV of calorimeter energy or total 
track pT within a cone of 0.4 = $\sqrt{\Delta\eta^{2}+\Delta\phi^{2}}$ 
of the track.








\section{Background}

The only background expected is from fakes, where a track has
a dE/dx which fluctuated high, or where overlapping tracks reconstruct
as a single track.  In order to estimate the background, we use a
control sample at low momentum (15 $< |\vec{p}| <$ 35 GeV) to
calculate the expected fake rate, then multiply the fake rate
by the number of tracks which enter our sample in the signal
region $|\vec{p}| >$ 35 GeV to get the expected background.
The fake rate is calculated in the control sample 
by taking the ratio of the number
of tracks which pass the dE/dx cut in both the CTC and the SVX
to the total number of tracks.

Since the fake rates are extrapolated from low 
momentum to high momentum, we
plot the fake rate as a function of momentum and check that
it is flat.  We observe that the fake rate falls off below 
20 GeV/c, especially in the CTC.  This is because our control
sample is contaminated with kaons, which are still along the
relativistic rise on their dE/dx vs. p curve below 20 GeV/c 
and pull down the fake rate
in this region in momentum.  For the muon triggered data sample, 
we take the control sample to be tracks which lie in the
momentum region 20 $< |\vec{p}| <$ 35 GeV and for the
 $E_{T}\hspace{-0.17in}/\hspace{0.1in}$ sample and the
electron sample our control region is 25 $< |\vec{p}| <$ 35 GeV.
 
\section{Results}
The results of each search are tabulated below. 


 \begin{table}
 \caption{Strong Search}
 \begin{tabular}{lrr} 
 Sample & Bkgd. & Obs. \\ 
 \tableline 
 Muon Trigger& 12 $\pm$ 2 & 12 \\ 
$E_{T}\hspace{-0.17in}/\hspace{0.1in}$ Trigger & 63 $\pm$ 9 & 45 \\
Muon Trigger ($\beta\gamma < $0.70)& 2.5 $\pm$ 0.8 & 2 \\
\end{tabular}
\end{table}

 \begin{table}
 \caption{Weak Search (Includes Isolation Cut}
 \begin{tabular}{lrr} 
 Sample &Bkgd.&Obs.\\  
 \tableline 
Muon Trigger& 0.85 $\pm$ 0.25 & 0\\
$E_{T}\hspace{-0.17in}/\hspace{0.1in}$ Trigger &4.0 $\pm$ 2.8 &1\\
Electron Trigger & 0.72 $\pm$ 0.54 & 0\\
\end{tabular}
\end{table}

\begin{figure}[ht]      

\centerline{\epsfxsize 2.3 truein \epsfbox{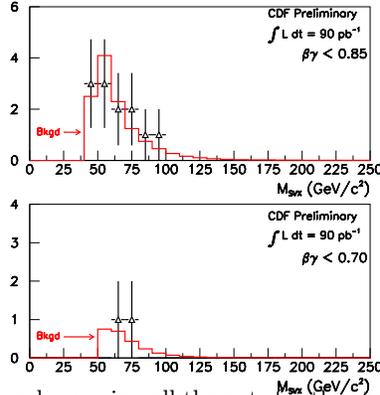}}
\vskip -.2 cm
\caption[]{
\label{Mass_dist_utrig}
\small Observed M$_{SVX}$ distribution for tracks passing all the cuts
for the strong search in the muon triggered data sample.  The solid
curves are the expected background distributions.}
\end{figure}

\begin{figure}[ht]      

\centerline{\epsfxsize 1.8 truein \epsfbox{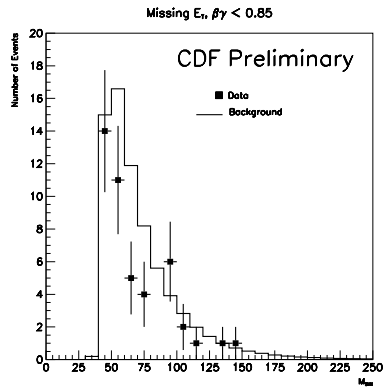}}
\vskip -.2 cm
\caption[]{
\label{Mass_dist_mettrig}
\small Observed M$_{SVX}$ distribution for tracks passing all the cuts
for the strong search in the $E_{T}\hspace{-0.17in}/\hspace{0.1in}$ 
triggered data sample.  The solid
curve is the distribution of the 63 $\pm$ 9 expected background events.}
\end{figure}

We calculate the expected mass distributions from the control sample.
To do this, we take the momentum of each track and calculate
what its dE/dx would be if it came from a particle with mass M,
and then use the control sample to find the probability that
the track fakes that dE/dx.  We do this for the entire range
in M and sum over all tracks.  The data and expected 
background distributions
are shown in Figures~\ref{Mass_dist_utrig} and ~\ref{Mass_dist_mettrig}.
No excess over background is observed.

\section{Efficiencies and Systematics}

For the strong search, the efficiency depends on the 
quark charge due to fragmentation
effects. It increases from
1.5 - 3\% for q=2/3 and 0.8-1.5\% for q=1/3 in the mass
range 100-240 GeV/c$^{2}$.  The largest systematic
uncertainties come from modeling interactions in the
calorimeter.  This gives an uncertainty of 20\% for 
q=1/3 and 13\% for q=2/3.  Systematic uncertainties
from luminosity (7.5\%) and from the choice of PDF (7\%)
are also significant.  The total systematic uncertainty
is 23\% and 17\% for q=1/3 and q=2/3 respectively.

For the weak search, the total efficiency is 3\%
for Drell-Yan slepton production, to which only the
muon trigger is sensitive.  Once slepton production
from cascade decays are included, the efficiency for
finding sleptons using only the muon triggered data
sample increases
to 6\%.  The total efficiency after including cascades
and all three triggers is 8\%.  The largest systematics
on this efficiencies come from the luminosity (7.5\%).
Other systematics include track quality cuts (4.9\%)
and choice of PDF (5.5\%). Once cascades are included,
the $E_{T}\hspace{-0.17in}/\hspace{0.1in}$ trigger
and initial/final state radiation (ISR/FSR) become
significant systematics.

\section{Limits}

\begin{figure}[ht]      

\centerline{\epsfxsize 2.3 truein \epsfbox{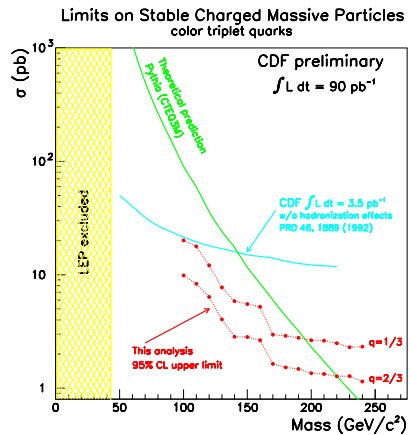}}
\vskip -.2 cm
\caption[]{
\label{hsq_limits}
\small Limits on heavy stable quarks derived from results of the strong
search.}
\end{figure}

Figure~\ref{hsq_limits} shows the cross-section limits 
we derive for heavy stable quarks
from the results of our strong search.  From comparison
with the theoretical prediction, we conclude M $>$ 195 GeV/c$^{2}$
for q=1/3 and M $>$ 220 GeV/c$^{2}$ for q=2/3.  These limits
are the most stringent limits from a direct search to date.

\begin{figure}[ht]      

\centerline{\epsfxsize 2.3 truein \epsfbox{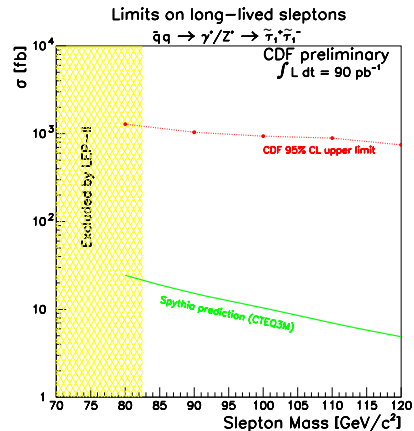}}
\vskip -.2 cm
\caption[]{
\label{slepton_limits}
\small Limits on long-lived Drell-Yan sleptons derived from 
results of the strong search.}
\end{figure}

Figure~\ref{slepton_limits} shows the cross-section limits derived for 
Drell-Yan production of long-lived sleptons in a GMSB
scenario with the stau as the Next-to-Lightest Susy Particle (NLSP).    
We are over an order of magnitude
away from being sensitive to the theoretical prediction.
When all sparticle production
modes with cascade decays to a slepton are included, 
the efficiency goes up, bringing
our limit down and the theoretically predicted
slepton production cross section increases.  This limit was derived
for one model point with three slepton co-NLSP's (N$_{5}$=3,
M/$\Lambda$=3, tan$\beta$=3, $\mu>$ 0) and found to be 
$\sigma_{95\%}\approx 550 fb$, still nearly an order of
magnitude away from the predicted 80 fb.

\section{Prospects for RunII}

RunII at the Tevatron will have an increased factor of 20
in luminosity and CDF will have a number of detector improvements.
The upgraded detector will have double-sided silicon with
stereo strips for z measurement, 7 silicon layers instead
of 4, and smaller cell size in the new tracking chamber.  This
smaller cell size will cut down on the background from overlapping
tracks.  Additionally, the now approved Time of Flight system
with a resolution of approximately 0.1ns will greatly increase our
region of sensitivity in momentum from $\approx$ 85 GeV/c to
$\approx$ 175 GeV/c for a 100 GeV/c$^{2}$ particle. 

\section{Conclusions}

We have searched for heavy stable particles using 90$pb^{-1}$
of Run1B data at CDF.  No excess of events over background was observed.
We set preliminary limits on heavy stable quarks at M $>$ 195 GeV/c$^{2}$ and
M $>$ 220 GeV/c$^{2}$ for q=1/3 and q=2/3 respectively.  
Using GMSB with three slepton co-NLSP's 
as a reference model, we set a limit on slepton production
at $\sigma_{95\%} < 550 fb$.  This limit is a factor of 7 away
from the theoretical prediction.  With increased luminosity and
detector upgrades in RunII we expect to have  
sensitivity to slepton production in GMSB models.

\section{Acknowledgements}
I am grateful to D. Stuart for
introducing me to this interesting topic, and
for his guidance throughout this analysis.
I would also like to thank M. Shapiro for her
advice and support.

\end{document}